\documentstyle[multicol,aps,epsf]{revtex} 

\tighten       

\begin{document}

\title{Stationary states and phase diagram for a model of the Gunn
 effect under realistic boundary conditions}

\author{ G. Gomila and J. M. Rub\'{\i}}
\address{Department de F\'{\i}sica Fonamental \\ 
	Universitat de Barcelona\\
 Diagonal 647, 08028 Barcelona, Spain}

\author{I. R. Cantalapiedra}
\address{Departament de F\'{\i}sica Aplicada\\
Universitat Polit\'{e}cnica de Catalunya\\
Gregorio Mara\~{n}\'{o}n 44, 08028 Barcelona, Spain}

\author{L. L. Bonilla}
\address{Escuela Polit\'{e}cnica Superior\\
Universidad Carlos III de Madrid\\
Butarque 15, 28911 Legan\'{e}s, Spain}

\date{\today}
\maketitle
\parskip 2ex

\begin{abstract}
A general formulation of boundary conditions for semiconductor-metal 
contacts follows from a phenomenological procedure sketched here.
The resulting boundary conditions, which incorporate only physically
well-defined parameters, are used to study the classical unipolar 
drift-diffusion model for the Gunn effect. The analysis of its stationary 
solutions reveals the presence of bistability and hysteresis for a certain 
range of contact parameters. Several types of Gunn effect are predicted to 
occur in the model, when no stable stationary solution exists,
depending on the value of the parameters of the
injecting contact appearing in the boundary condition. In this way, the
critical role played by contacts in the Gunn effect is clearly stablished.
\end{abstract}

\draft{PACS: 05.45+b, 72.20.Ht, 85.30.Fg, 05.70.Ln} 

\begin{multicols}{2}
\narrowtext

\section{Introduction}
The Gunn effect is an ubiquitous phenomenon in many semiconductor 
samples presenting negative differential resistance and subject to
voltage bias conditions \cite{Gunn,kro72,Grubin1,Grubin2}. In a nutshell, 
the negative differential 
resistance makes it possible the existence of a variety of pulses 
and wavefronts, which may be stabilized by the bias condition. Then 
a periodic shedding of waves by the injecting contact results in the 
periodic oscillation of the current through an external circuit, which 
constitutes the signature of the Gunn effect. There is a vast literature 
on this topic, despite of which different basic questions concerning 
the Gunn effect remain poorly understood. Paramount among these, there 
are the questions about which are the correct boundary conditions and, 
given these, how to describe all the stages of the Gunn oscillation.
The lack of a precise formulation of the boundary conditions imposed
by contacts on semiconductors and of a simple analytic treatment
to analyze the Gunn oscillations, has not allowed to clarify 
in a precise way the role played by contacts in the
Gunn effect. It is worth noting that clarifying this point
would open, for instance, the possibility
of extracting information about the contacts from the analysis of
the Gunn oscillations themselves, a subject of considerable interest
for applied researchers.

Recently progress has been made towards answering reasonably these two 
questions. On the one hand, ideas from Irreversible Thermodynamics 
\cite{Degroot} have been used to derive satisfactory boundary conditions 
for metal-semiconductor and other contacts in a general 
way \cite{Noltros1,Noltros2,Noltros3}. Previously the 
usual boundary conditions used in drift-diffusion semiconductor models 
were: (i) periodic,\cite{nakamura} (ii) charge neutrality,\cite{szmolyan} 
(iii) fixed field,\cite{Grubin1,Grubin2} and (iv) control current-field 
characteristics of the contact \cite{kro68} plus phenomenological 
assumptions such as the ``contact length'' \cite{Grubin1}. As boundary 
conditions (b.c.) for a semiconductor presenting the Gunn effect, 
these conditions rank from clearly wrong (no current oscillation appears 
if the b.c. are periodic) to unsatisfactory because of their ad hoc 
character. Thus even when numerical simulations display the Gunn effect, 
the question of whether these results describe the real physical system
 where different contacts are present, is usually raised.
In this paper, we shall present a simple derivation of appropriate 
b.c. for an ideal
metal-semiconductor (MS) contact and use them to analyze the Gunn effect 
in Kroemer's model for bulk n-GaAs. Our description makes it clear which
part of the derivation follows from general principles and which part 
includes the input from the physics of contacts. 

Concerning asymptotic descriptions of the Gunn effect which delve deeper 
than just numerical simulations of drift-diffusion models, some progress
has been made recently \cite{HB,BKMV,BHHKV,univ}. 
A detailed treatment of this topic can be found in Ref.\cite{companion}.

The rest of the paper is as follows. In Section II we present our derivation
of b.c. for ideal MS contacts and briefly discuss some other possibilities.
Kroemer's model and its stationary solutions for these b.c. are analyzed 
in Section III. It is found that bistability between stationary solutions
is possible for certain bias ranges depending on the values of certain
dimensionless contact parameters $i_0$, {\bf $\alpha_{0}$},
 which are a combination of its 
effective density of states, barrier height, Richardson's
constant, doping and temperature. Different types of Gunn effect, namely,
charge monopoles (moving charge
accumulation and depletion layers), and charge dipoles,
(high and low field solitary waves), are
predicted to appear depending on these contact parameters,
when no stable stationary solution exists.
In Section V we discuss our results, whereas Appendix A is devoted to
technical matters related to the main text.

\section{Boundary conditions}
The aim of this section is to present a systematic procedure to
derive b.c. at semiconductor contacts, established in previous
works \cite{Noltros1,Noltros2,Noltros3}. As a general rule,
the method applies for
non-degenerate semiconductors under moderate
temperatures, that is, when thermionic emission is the dominant
transport process at the contact. Hence
several contacts of interest, like ideal and non-ideal
metal-semiconductor (MS) contacts or any type of heterojunction
contact, can be modeled. Depending on the material parameters
both limiting as well as ohmic contacts may then be described.
It is worth noting that a precise modeling of this type of contacts
may help to clarify the role played by other types of contacts
used in 
semiconductor systems, e.g. those in
which thermionic-field or field emission processes
dominate \cite{Rideuot,Brillson,Monch}, for
which a precise description, in the sense of the present paper,
is not yet available.
For the sake of clarity, the method will be presented along with 
its application to the case of an ideal MS contact. Other contacts
have been considered in previous papers \cite{Noltros1,Noltros2,Noltros3}.

Let us consider an ideal MS contact. Due to the presence of the 
contact, the magnitudes describing the physical properties of the
system, e.g electron density, electric field, electron energy, \ldots,
may be discontinuous at that point. In addition, singular contributions
localized at the contact itself, e.g. electron density at interface
states (when they are present), may also occur. As a consequence, a 
given physical magnitude,
$d(\tilde{x},\tilde{t})$, can be decomposed as follows
\begin{equation}
d(\tilde{x},\tilde{t})=d_{n}(\tilde{x},\tilde{t}) \theta(\tilde{x}) +
 d_{m}(\tilde{x},\tilde{t}) 
\theta(-\tilde{x}) + d_{s}(\tilde{t}) \delta (\tilde{x}) ,
\label{descom}
\end{equation}
where $d_{n}$, $d_{m}$, $d_{s}$ refer to the values in the semiconductor ($n$),
metal ($m$), and surface ($s$) parts , respectively (when no singular
contribution is present, $d_{s}$ vanishes). Moreover, $\theta(\tilde{x})$ is 
Heaviside's unit step function and $\delta(\tilde{x})$ Dirac's delta 
function. They are introduced
in order to represent the discontinuity across the contact and the singular
contributions, respectively. In writing Eq.(\ref{descom}), a one 
dimensional description
of the system has been assumed, with the contact being located at 
$\tilde{x}=0$ and
the metal [semiconductor] on its left [right]. By means of this type of
decomposition, b.c. can be systematically derived.

Our procedure consists of two steps: a) the identification of the relevant
magnitudes describing the transport processes through the contact,
 and b) the derivation of precise laws describing
such processes, which relate the relevant magnitudes at the contact and
which constitute the desired b.c..
For the first step, use will be made of a phenomenological formulation
of transport through semiconductor junctions\cite{Noltros1} while for the 
second of Shockley-Read-Hall (SRH) statistics \cite{Shockley,Hall}.

Let us consider a given magnitude, $d(\tilde{x},\tilde{t})$, satisfying a 
standard balance equation of the form \cite{Degroot}
\begin{equation}
\frac{\partial d(\tilde{x},\tilde{t})}{\partial \tilde{t}} + 
\frac{\partial J_{d}(\tilde{x},\tilde{t})}{\partial \tilde{x}}=
\sigma_{d}(\tilde{x},\tilde{t}) ,
\label{balance1}
\end{equation}
where $J_{d}(\tilde{x},\tilde{t})$ and $\sigma_{d}(\tilde{x},\tilde{t})$ 
refer to the current and
net rate production associated to the magnitude $d$, respectively. It is 
not difficult
to show that if similar balance equations were to be satisfied on each
side of the junction and if surface fluxes only exist along the interface
(what in a one dimensional description means $J_{d,s}(\tilde{t})=0$), then
the following balance equation should be satisfied at the contact, 
\cite{Bedeaux75}
\begin{equation}
\frac{\partial d_{s}(\tilde{t})}{\partial \tilde{t}} + [J_{d,n}(0,\tilde{t})-
J_{d,m}(0,\tilde{t})] = \sigma_{d,s}(\tilde{t}) .
\label{balance}
\end{equation}
Now, we can proceed to calculate the net rate of entropy production at
the contact, which will allow us to identify the relevant magnitudes describing
the transport processes trough the contact.
To begin with, we consider the balance equation for the total energy of the
system. As this is a conserved quantity, we simply have
\begin{equation}
\frac{\partial e_{s}(\tilde{t})}{\partial\tilde{t}} + 
[J_{e,n}(0,\tilde{t})-J_{e,m}(0,\tilde{t})] = 0 .
\label{balancee}
\end{equation}
As for an ideal MS contact no interface states are present, and hence no net
charge or mass is accumulated at the contact, the total energy at the
contact coincides with the surface internal energy, $u_{s}=e_{s}$ 
\cite{Bedeaux75}.
 Hence, the
balance of internal energy is described directly through Eq.(\ref{balancee})
or alternatively through
\begin{equation}
\frac{\partial u_{s}(\tilde{t})}{\partial \tilde{t}} + 
[J_{u,n}(0,\tilde{t})-J_{u,m}(0,\tilde{t})] = 
                                   \sigma_{u,s}(\tilde{t}) ,
\label{balanceu}
\end{equation}
with $\sigma_{u,s}(\tilde{t}) = [J_{u,n}(0,\tilde{t})-J_{e,n}(0,\tilde{t})] -
[J_{u,m}(0,\tilde{t})-J_{e,m}(0,\tilde{t})]$. In the previous expression, we 
have introduced 
explicitly the flux of internal energy (equivalent to the heat flux), which 
is in general different
from the flux of total energy. Furthermore the Gibbs equation for an ideal 
contact is \cite{Bedeaux75} $T ds_{s} = du_{s}$ (no interface states
are present), where $s_{s}$ is the surface entropy and $T$ the
temperature. By assuming the contact to be in local equilibrium, one 
then has $T \partial s_{s} (\tilde{t}) /\partial \tilde{t} =  
\partial u_{s}(\tilde{t})/\partial \tilde{t}$, which, after 
using Eq.(\ref{balanceu}), gives rise to  the balance equation for the entropy 
\begin{equation}
\frac{\partial s_{s}(\tilde{t})}{\partial \tilde{t}} + 
[J_{s,n}(0,\tilde{t})-J_{s,m}(0,\tilde{t})] = 
                                   \sigma_{s,s}(\tilde{t}) ,
\label{balances}
\end{equation}
with the entropy production given by
\begin{eqnarray}
\sigma_{s,s}(\tilde{t})&=& [J_{s,n}(0,\tilde{t}) - 
\frac{1}{T} J_{e,n}(0,\tilde{t})] 
                   \nonumber \\
                       & &-[J_{s,m}(0,\tilde{t}) - 
                       \frac{1}{T} J_{e,m}(0,\tilde{t})] .
\label{sigmas}
\end{eqnarray}
A more explicit expression for $\sigma_{s,s}$ is obtained once the 
bulk expressions
for the fluxes are introduced on the r.h.s. of Eq.(\ref{sigmas}). These
expressions can be found elsewhere \cite{Degroot,Noltros1}. One has,
\begin{equation}
J_{s,a}(0,\tilde{t}) = \frac{1}{T} J_{e,a}(0,\tilde{t}) - 
\frac{1}{T} E_{F,a}(0,\tilde{t}) J_{a}(0,\tilde{t}) \mbox{;} \hspace{0.25 cm}
 a=m,n
\end{equation}
where $E_{F,a}$ refers to the electron Fermi level (or chemical potential) and
$J_{a}$ to the electron number density current. Substituting into 
Eq.(\ref{sigmas}), we simply have
\begin{eqnarray}
\sigma_{s,s}(\tilde{t}) &=& - \frac{1}{T} [E_{F,n}(0,\tilde{t}) 
J_{n}(0,\tilde{t}) - E_{F,m}(0,\tilde{t}) J_{m}(0,\tilde{t})] \nonumber \\
                        &=&  - \frac{1}{T} [E_{F,n}(0,\tilde{t}) - 
                        E_{F,m}(0,\tilde{t})] J_{n}(0,\tilde{t}) ,
\end{eqnarray}
where in the second line use has been made of the continuity of the
electron number density current at an ideal MS contact (this continuity
follows from the corresponding balance equation for the electron number
density by imposing that no carriers are accumulated ($n_{s}(\tilde{t})=0$) nor
created ($\sigma_{n,s}(\tilde{t})=0$) at the contact). The final expression to
be used in what follows, is obtained by introducing the electron quasi-Fermi
levels, $F_{a}(0,\tilde{t})=E_{F,a}(0,\tilde{t}) - e V_{a}(0,\tilde{t})$. 
Here $V_{a}(0,\tilde{t})$ is the electric potential (which is continuous 
through an abrupt junction) and $e>0$ is minus the charge of the electron. 
We then arrive at the desired expression, 
\begin{equation}
\sigma_{s,s}(\tilde{t}) = - \frac{1}{T} [F_{n}(0,\tilde{t}) - 
F_{m}(0,\tilde{t})] J_{n}(0,\tilde{t}) .
\label{finals}
\end{equation}
Eq.\ (\ref{finals}) shows directly that the relevant magnitudes describing
an ideal MS contact are the electron flux (electron current density divided
by $e$), $J_{n}(0,\tilde{t})$, and the discontinuity in the electron 
quasi-Fermi levels, $(F_{n}(0,\tilde{t}) - F_{m}(0,\tilde{t}))$, which 
plays the role of ``thermodynamic force'', \cite{Degroot}. Both flux and
force vanish at equilibrium, and we assume (in accordance with the basic
tenets of Irreversible Thermodynamics \cite{Degroot}) that there is a 
relation between them. When the fundamental relation between flux and 
force is specified, this relation is exactly the sought-after boundary 
condition at the contact. 

The relation between $J_{n}(0,\tilde{t})$ and $(F_{n}(0,\tilde{t}) - 
F_{m}(0,\tilde{t}))$ should involve more information about the physics 
of the contact. First of all, let us notice that the entropy production 
(\ref{finals}) is formally equivalent to the expression corresponding
to generation-recombination processes \cite{Vanvliet} (or in general, to
any activated process, such as unimolecular chemical reactions \cite{Degroot} 
or surface adsorption \cite{Ignasi92}), provided $J_{n}(0,\tilde{t})$ 
is identified with the net rate of the process. From this comparison we 
then conclude that the transport through an ideal MS contact may be 
described as an elementary kinetic process of the form: $q_{m} 
\rightleftharpoons q_{n}$, where $q_{n}$ and $q_{m}$ represent the 
carriers in the semiconductor and metal, respectively, with
the {\em net rate of the process} being equal to $J_{n}(0, \tilde{t})$. The 
kinetics of 
such a process can be described, for instance, by means of SRH statistics.
As it is well known, this description relates the kinetic rate of
the process, in our case $J_{n}(0, \tilde{t})$, to the affinity, in our case
the difference in quasi-Fermi levels, in agreement
with our former general treatment.
As it is shown in Appendix \ref{appendixA}, we obtain the following relation 
between the net rate of the process, $J_{n}(0,\tilde{t})$, and the 
jump of the quasi-Fermi level at the MS contact, $(F_{n}(0,\tilde{t}) - 
F_{m}(0,\tilde{t}))$:
\begin{equation}
J_{n}(0,\tilde{t}) = \lambda_{0} e^{- e \beta \phi_{b}^{0}}
\left( 1 - e^{ \beta (F_{n}(0,\tilde{t}) - F_{m}(0,\tilde{t}))} \right) .
\label{TE}
\end{equation}
Here $\lambda_{0}$ is a positive constant [see (\ref{lambda}) in Appendix 
\ref{appendixA}], 
and $e \phi_{b}^{0} = E_{C} (0,\tilde{t}) - F_{m}(0,\tilde{t})$ is
the contact barrier height. $E_ C(0,\tilde{t}) = E_C^0 - e V_n(0,\tilde{t})$,
is the electron energy, with $E_C^0$ the bottom of the  semiconductor 
conduction band and $V_n$ the 
 electric potential at the semiconductor surface. Moreover,
 $\beta=(k T)^{-1}$, where $k$ is the Boltzmann constant. Eq.(\ref{TE}) 
is the sought b.c. for the ideal MS contact. It can be written more 
explicitly by using the expression $\tilde{n} = N_{C}  e^{ \beta (F_{n} 
- E_{C})}$, which holds for non-degenerate semiconductors with
$\tilde{n}$ being the semiconductor electron number density 
and $N_{C}$ the effective density of states. We then have,
\begin{equation}
J_{n}(0,\tilde{t}) = \frac{\lambda_{0}}{N_{C}} \left( N_{C} 
e^{- e \beta \phi_{b}^{0}} - \tilde{n}(0,\tilde{t}) \right) \,.
\label{sze}
\end{equation}

In the non-degenerate case, $\lambda_{0}$ only depends on $T$ (see Appendix 
\ref{appendixA}). This result ends the derivation
of the b.c. for ideal MS contacts which we will use for the rest of this 
paper. It is worth emphasizing at this point that, as mentioned at 
the beginning of this section, the procedure we have sketched here for
the ideal MS contact, can be applied to several other types of contacts.
These include non-ideal MS contacts with the presence of interface states 
\cite{Noltros2} and unipolar or bipolar heterojunction contacts with or 
without interface states \cite{Noltros2}. Moreover, by adding a few 
assumptions one can handle non-abrupt contacts \cite{Noltros3}.

Note that for a MS contact
located at $\tilde{x}=\tilde{L}$, that is, with the metal [semiconductor]
 on the right [left] hand side of the
contact, the corresponding b.c. is
\begin{equation}
J_{n}(\tilde{L},\tilde{t}) = -\frac{\lambda_{L}}{N_{C}} \left( N_{C} 
e^{- e \beta \phi_{b}^{L}} - \tilde{n}(\tilde{L},\tilde{t}) \right) .
\label{szeL}
\end{equation}

We can compare our  result, Eq.(\ref{sze}) for the ideal MS contact 
with the corresponding one reported in \cite{Sze2} [see also p.\ 261 
of Ref.\cite{Sze}]. Then we can identify  
$\lambda_{i}= A_{i} T^{2}/e$, $i=0,L$, where $A_{i}$ is the 
Richardson constant
for the semiconductor in contact with the metal located at $i=0,\tilde{L}$.
 Theoretically, $A_{i}$, and hence 
$\lambda_{i}$, would depend only on the given 
semiconductor but not on the metal \cite{Sze}. However, in practice $A_{i}$ is
taken as a phenomenological parameter and it can not only depend on the metal
but also on the preparation procedures \cite{Eftekary}.
On the other hand, basic energetic arguments lead immediately to the following 
rule 
for the contact barrier height \cite{Sze,Roderick}: $\phi_{b}^{i} = 
\phi_{M}^{i} - \chi$. Here $\phi_{M}^{i}$ is the work function of the metal 
in the MS contact located at $\tilde{x}=i$ and $\chi$ is the semiconductor 
electron affinity. For covalent semiconductors, the validity of this rule 
has been put under question
for the last five decades \cite{Rideuot,Brillson}. However, recently
it has been shown that even for this type of semiconductors, if accurately
growth materials are used, good agreement is obtained with this simple
rule \cite{Vituro1,Vituro2}. Notice, that when such materials are not
used, as very often happens, the contact formed turns out to be non-ideal. 
This
is so because it is difficult to avoid that very thin insulating layers 
and/or interface states may be present at the contact \cite{Brillson,Monch}.
Hence, a non-ideal description of the contact should be used, with 
for instance, a bias dependent relation for the barrier height which 
would include the effects 
of the insulating layer and of the interface states. Such contacts have been
described in detail in Ref. \cite{Noltros2,Noltros3}. In particular,
it has been shown that for this non-ideal contacts there is no simple 
description using 
directly equations such as (\ref{TE}) or (\ref{sze}), for contact 
non-stationary effects induced by interface sates (not present for 
ideal MS contacts) introduce additional terms not present in Eq.(\ref{TE}). 
These terms could be of importance when describing naturally 
non-stationary phenomena such as the Gunn effect with non-ideal contacts,
and will be considered in future works.

Eqs.~(\ref{sze}) and (\ref{szeL}) can be rewritten in terms of the
electric field by using the Poisson equation to eliminate the electron 
density from them:
\begin{equation}
\frac{\partial \tilde{E}}{\partial \tilde{x}} (i,\tilde{t}) = 
\frac{e}{\varepsilon} \left( \tilde{n}_{0} - N_{C} e^{-e \beta \phi_{b}^{i}}
   \pm \frac{J_{n}(i,\tilde{t})}{\frac{\lambda_{i}}{N_{C}}} \right)\,.
\label{bc}
\end{equation}
Here the upper [lower] sign holds for $i=0$ [$i=\tilde{L}$]; $\varepsilon$ and 
$\tilde{n}_0$ are the bulk semiconductor permitivity, and its doping, 
respectively.

We have thus shown how our procedure allows us to derive explicit and
precise expressions for the b.c. imposed by a given contact.
A first important consequence of this method can be drawn directly 
from Eq.~(\ref{bc}), which is simply a relation between the normal {\em
derivative} of the electric field and the current density at the contact. 
Examining our derivation shows that this result is simply 
a consequence of the use of kinetic models to describe the exchange of 
carriers through the contact. Hence, one should expect that b.c. derived in
this way, will result in relationships between the normal derivative of the 
electric field and the current density at the contact. It is easily seen
that (if diffusion effects can be neglected) our b.c. can be transformed 
into a Kroemer's type contact current-field control characteristics
\cite{kro68} (see next section). However, unlike previous models following 
Kroemer's approach \cite{Grubin1,Grubin2,HB,Shaw}, our control 
characteristics is the result of a physically precise derivation, and 
therefore only parameters which are physically well-defined appear in it. 
In particular, to use our control characteristics we do not have to invoke 
ad hoc assumptions involving new parameters such as the contact length 
\cite{Grubin2,Shaw}.

To facilitate the analysis in the rest of the paper, it is convenient to 
rewrite Eq.~(\ref{bc}) in dimensionless units. This greatly reduces the 
number of relevant parameters. Our dimensionless electric field 
$\cal{E}$, electron densities $n$, current densities $j(x,t)$, time 
$t$ and position $x$ are measured in units of $\tilde{E}_0$, 
$\tilde{n}_0$, $e \tilde{n}_0 \mu_0 \tilde{E}_0$, $\varepsilon/(e \mu_0 
\tilde{n}_0)$ and $\varepsilon \tilde{E}_0/(e \tilde{n}_0)$, 
respectively~\cite{BonillaPD91}. In these equations,
$\tilde{E}_0$ and $\mu_0$ are an electric field and the zero-field electron 
mobility typical of the processes occurring in the bulk of the semiconductor
(see below). Then Eq.~(\ref{bc}) becomes 
\begin{equation}
\frac{\partial \cal{E}}{\partial x}(i,t) = \alpha_{i} [ - i_{i} 
               \pm j(i,t) ] ,
\label{bc1}
\end{equation}
where we have defined
\begin{eqnarray}
\alpha_{i} &=& \frac{\mu_{0} \tilde{E}_{0}N_{C}}{\lambda_{i}} ,
\label{alfa} \\
i_{i}      &=& - \alpha_{i}^{-1} \left( 1 - 
            \frac{N_{C}}{\tilde{n}_{0}} e^{-e \beta \phi_{b}^{i}} \right).
\label{i}
\end{eqnarray}              
As mentioned before, here the upper [lower] sign refers to $i=0$  
[$i=L$]. It is worth noting that $\alpha_{i}$ is always a positive 
quantity (because $\lambda_{i}$ is) while $i_{i}$ does not have a 
definite sign: it depends basically on the value of the barrier height,
$\phi_{b}^{i}$, and on the doping value, $\tilde{n}_{0}$.  

It should be noted that there are important restrictions on the 
possible values of the contact current density which are due to the fact 
that in Eqs.~(\ref{sze}) and (\ref{szeL}) the electron density, $n$, is a 
positive quantity: 
$$0\le \tilde{n}(i,\tilde{t}) = N_{C} e^{- e \beta \phi_{b}^{i}} \pm 
\frac{N_{C}}{\lambda_{i}}\, J_{n}(i,\tilde{t}),$$
or in dimensionless units
\begin{equation}
 \pm j(i,t) < \alpha_{i}^{-1} + i_{i} \equiv j^{sat}_i. \label{jsat}
\end{equation}
[Eqs.~(\ref{alfa}) and (\ref{i}) imply that $\alpha_{i}^{-1} + i_{i}
=j_i^{sat}$ is always a positive quantity, equal to the maximum 
current density which the contact can provide]. These restrictions on 
the current are reminiscent of the rectifying properties of MS 
contacts. In practice, they 
only impose a real limitation for the case of true rectifying 
contacts (when one of the $j_i^{sat}$ is small). Otherwise, i.e., for large 
values of $j_i^{sat}$, an ohmic contact is obtained
which does not impose a real limitation on the current. 

In order to analyze the influence of the derived b.c.
on the Gunn instability, we will assume a sample formed by a
certain semiconductor (able to display the Gunn effect) and by
two MS contacts implemented on it. The resulting b.c. are
\begin{eqnarray}
\frac{\partial \cal{E}}{\partial x }(0, t) & = & \alpha_{0} 
        [j(0,t) - i_{0}], \label{bc0} \\
\frac{\partial \cal{E}}{\partial x }(L,t) & = & - \alpha_{L} 
          [i_{L} + j(L,t) ] .\label{bcl}
\end{eqnarray}
As discussed above, for a given semiconductor the values of
the contact parameters may vary somewhat depending on the metal
used in the contact and on the preparation procedures. For instance
$\alpha_{i}$ may vary two orders of magnitude, from about $0.3$ to $33.4$,
if we use the experimental values of Richardson's constant for GaAs 
reported in \cite{Eftekary}. Similarly, the values of $i_{i}$ may also 
span two orders of magnitude, from about $0.03$ to $4.01$, due to the 
variation of $\alpha_{i}$, if we fix the barrier height, $\phi_{b} 
\approx 0.2$ V (corresponding to Al \cite{Vituro1}), and the donor
density is $10^{14}$ cm$^{-3}$. Thus there is a rather
wide range of parameter values for the contacts, corresponding to
a large variety of situations which will be described in this paper.

Lastly, the applied bias $V$, defined as $e V(t) = 
\hat{F}_{m}(L,t) - \hat{F}_{m}(0,t)$, can be 
expressed as follows: $V(t) = \int_{0}^{L} {\cal{E}} (x,t) \mbox{d}x + 
(\hat{\phi}_{b}^{0} - \hat{\phi}_{b}^{L})$. In the previous
expressions, $V$ and $\hat{\phi}_{b}^{i}$ are
in units of $\varepsilon \tilde{E}_{0}^{2}/ e \tilde{n}_{0}$ and
$\hat{F}_{m}$ in units of $\varepsilon \tilde{E}_{0}^{2}/ \tilde{n}_{0}$.

Very frequently the analysis of the Gunn effect under dc
voltage bias is carried out by using the opposite sign for the electric
field: $E = - \cal{E}$. Then the b.c. become
\begin{eqnarray}
\frac{\partial E}{\partial x}(0,t) & = & 
\alpha_{0} [i_{0} - j(0,t) ],  \label{bc01} \\
\frac{\partial E}{\partial x}(L,t) & = & 
\alpha_{L} [i_{L} + j(L,t) ]  \label{bcl1}
\end{eqnarray}
With these definitions,
the dc voltage is $V = -\int_{0}^{L} E \mbox{d}x  
-(\hat{\phi}_{b}^{L} - \hat{\phi}_{b}^{0})$. Instead of working with 
the voltage $V$, it is more convenient to use the average electric 
field on the 
semiconductor sample, $\phi = L^{-1}\int_0^L E(x,t) $d$x$, 
which is equal to $\phi = \frac{1}{L}[-V + (\hat{\phi}_{b}^{0}
-\hat{\phi}_{b}^{L})]$. In what follows, negative voltages, $V<0$, will
be considered such as $\phi>0$. With these conventions, the carriers go 
from the cathode (injecting contact) at $x=0$ to the anode 
(receiving contact) at $x=L$.

\section{Kroemer's model and its stationary states}

\subsection{Kroemer's model}

The unipolar drift-diffusion model for the Gunn effect proposed 
by Kroemer \cite{kro72,Kroemer}, is generally accepted to provide a
rather complete description of the main features of this effect.
Yet it is simple enough to allow very detailed asymptotic analysis;
other important models such as B\"uttiker and Thomas's \cite{buttiker} 
incorporate more detailed physics but their study is technically more 
demanding. In the dimensionless units described above, Kroemer's model is
\begin{eqnarray}
{\partial E\over\partial t} + v(E)\, \left( {\partial E\over\partial x} 
+ 1\right) - \delta {\partial^{2} E\over\partial x^{2}} = J,\label{eq}\\
{1\over L}\, \int_0^L E(x,t)\, dx = \phi. \label{bias}
\end{eqnarray}
Eq.~(\ref{eq}) is Amp\`ere's law which
says that the sum of displacement current and drift-diffusion current is
equal to the total current density $J(t)$. It can be obtained by 
differentiating the Poisson equation, $\partial E/\partial x = n-1$, with
respect to time, substituting the charge continuity equation
$\partial n/\partial t + \partial j(x,t)/\partial x = 0$ [the electron
current density is of the drift-diffusion type: $j(x,t) = n v(E) - 
\delta\, \partial n/\partial x$], and then integrating the result with 
respect to $x$. The electron velocity is assumed to be 
N-shaped and for specific calculations we shall use~\cite{Kroemer}
\begin{equation}
v(E) = E\, {1 + B E^{4}\over 1 + E^{4}},\label{v}
\end{equation}
(it has a maximum $v_M>0$ at $E_M>0$ followed by a minimum $0<v_m <
v_M$ at $E_m>E_M$) and the electron difusivity $\delta$ to be constant. 
The results using other curves having the same shape are similar. 
If $v(E)$ does not reach a minimum but saturates instead as $E\to\infty$, 
not all the monopole and dipole waves which we have found occur. Thus we 
have chosen the velocity curve that yields the richest dynamical behavior. 
The behavior of Kroemer's model with saturating velocity will be 
commented upon in the discussion. The dc 
bias $\phi$ is the average electric field on the semiconductor sample. 
Eqs.~(\ref{eq})-(\ref{bias}) need to be solved with an appropriate initial 
field profile $E(x,0)$ and subject to the following mixed boundary conditions
resulting from substituting $j(x,t) = J(t) - \partial E/\partial t$ [from
(\ref{eq})] into (\ref{bc01})-(\ref{bcl1}):
\begin{eqnarray}
\frac{\partial E}{\partial x}(0,t) & = & \alpha_{0} \left(i_{0} - J(t) + 
\frac{\partial E}{\partial t}(0,t)\right)\,, \label{bc02} \\
\frac{\partial E}{\partial x}(L,t) & = & \alpha_{L}  \left(i_{L} + J(t) - 
\frac{\partial E}{\partial t}(L,t)\right)\, . \label{bcl2}
\end{eqnarray}

In what follows, $i_{i}$ will be assumed to be positive because 
the physically interesting  phenomena (including the 
usual Gunn effect mediated by high field domains) are observed
for these values of $i_{i}$, as will be seen in the following sections.

For typical n-GaAs data, $\delta\ll 1$ and $L\gg 1$ \cite{HB}. In this 
limit, we shall find approximate solutions to the initial boundary value
problem (\ref{eq})-(\ref{bcl2}) for $E(x,t)$ and $J(t)$. 
Strictly speaking, the simple asymptotic description that follows holds 
in the limit $L\to\infty$, even when $\delta =O(1)$~\cite{univ}. Assuming 
$\delta\ll 1$ just simplifies the description of the traveling waves of 
electric field in the semiconductor through the use of characteristic 
equations and shock waves~\cite{HB,Knight,SIAM91}.

To take advantage of this limit, we will use the following rescaled
time and length, 
\begin{equation}
\epsilon = \frac{1}{L}\,,\quad s = {t\over L}\,,\quad
\quad y = {x\over L}\, .\label{slow}
\end{equation}
Then Eqs.~(\ref{eq})-(\ref{bias}) become 
\begin{eqnarray}
J - v(E) = \epsilon \left( {\partial E\over\partial s} + v(E)\, 
{\partial E\over\partial y} \right) - \delta\epsilon^2 {\partial^{2} 
E\over\partial y^{2}} \,,\label{s-eq}\\
\int_0^1 E(y,s)\, dy = \phi. \label{s-bias}
\end{eqnarray}

\subsection{Stationary states and its stability}
Before describing the Gunn effect in the present model, it is convenient
to discuss how to construct the stationary solutions of the model in the
limit $\epsilon\ll 1$ and $\delta\ll 1$. (In the case $\delta = O(1)$ the 
procedure is slightly more complicated and we shall omit the corresponding 
details; see \cite{BonillaPD91}. 
In this section we shall analyze the stationary states of 
Kroemer's model in 
n-GaAs~\cite{kro72,Kroemer} under dc voltage bias with the new 
boundary conditions (\ref{bc01})-(\ref{bcl1}). Our work is
based upon previous asymptotic and numerical studies of this and related
models~\cite{HB,BKMV,BHHKV,univ}. 

\begin{figure}     
\centerline{
    \epsfxsize=9 cm
    \epsffile{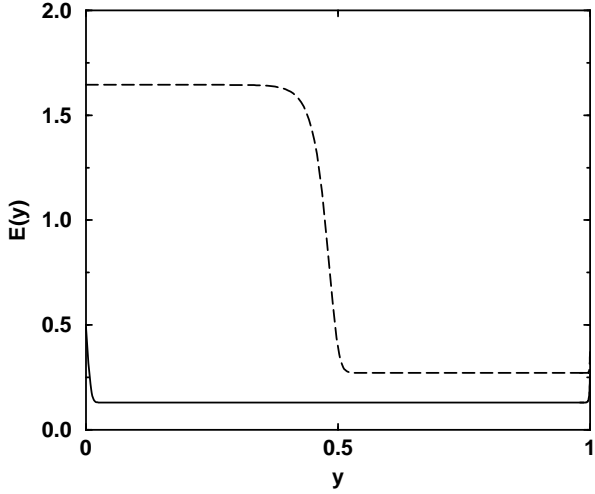}}
\caption{Stationary electric field profiles, showing the 
piecewise character of the solutions. The dashed line corresponds to a
step-like stationary solution. Narrow boundary layers are present at the 
contacts.}
\label{sprof}
\end{figure}

In this asymptotic limit, any stationary solution can be described
as composed of outer and inner solutions: the outer bulk solution is 
a piecewise constant field profile valid everywhere except for 
two narrow boundary layers located at the contacts and, for 
particular values of the current density, a 
narrow transition layer somewhere in the middle of the sample 
(see Fig.\ \ref{sprof} and explanation below). 
First of all, if we ignore inner solutions, 
the stationary state solves the equations 
\begin{eqnarray}
 v(E) - J = O(\epsilon), \nonumber\\
E = \phi + O(\epsilon).\label{ss1}
\end{eqnarray}
except for particular values of $J$ which will be specified below.
These equations result from retaining only order-one terms 
in Eqs.(\ref{s-eq}) and (\ref{s-bias}), and assuming $E(y) =$const.
Then for those values of $\phi$ such that the
outer solution (\ref{ss1}) is compatible with the boundary conditions,
we have $J=v(\phi) + O(\epsilon)$. 
Let us denote by $E_1(J)<E_2(J)<E_3(J)$ the three zeros of $v(E)-J$
[$E_2(J)$ is unstable]. Then the outer (bulk) field profile will be 
$E(y) =E_{i}$, $i=1,3$, depending on the value of the bias $\phi$.

At $y=0$ and $y=1$ there are boundary layers, which we will call {\em 
injecting and receiving layers}, respectively. $E(y;J)$, the field at the 
injecting boundary layer of width $O(\epsilon)$ at $y=0$, obeys [we ignore 
the $O(\delta)$ diffusive term]: 
\begin{eqnarray}
\epsilon {\partial E\over\partial y} = {J\over v(E)} - 1, \quad x>0,\label{ss2}\\
\epsilon {\partial E\over\partial y}(0;J) = \alpha_0 \, (i_0 - J). \label{ss3}
\end{eqnarray}
The analysis of Eqs.(\ref{ss2}) and (\ref{ss3}) is more easily carried
out if we express the derivative b.c. Eq.(\ref{ss3}) in terms of a b.c. 
for the electric field at the contact, $E(0;J)$  .
We can obtain $E(0;J)$ from (\ref{ss3}) by using (\ref{ss2}) to eliminate 
$\partial E(0;J)/\partial y$. The result is that $E=E(0;J)$ solves
\begin{eqnarray}
j_c(E) &=& J,\quad\quad\mbox{where} \label{ss4}\\
j_c(E) &=& {(1+\alpha_{0} i _{0})\, v(E)\over1+\alpha_{0}\, v(E)} 
\, . \label{ss5}
\end{eqnarray}
Notice that the contact curve $j_c(E)$ has the same extrema as the 
velocity curve $v(E)$ and saturates for high electric fields
to the value $j_{0}^{sat}$, defined in Eq.(\ref{jsat}). Kroemer's contact 
characteristic for 
shallow-barrier metal-semiconductor contacts presented in 
\cite{Kroemer}, corresponds to a particular case of our model in 
which the electrons in the metal are assumed to be in equilibrium 
with those of the semiconductor near the contact. For this case, one 
would take $\alpha_0 \to 0$ with $|\alpha_0 i_0 | < \infty$, so 
that $j_c(E)$ would be then proportional to $v(E)$. In 
contradistinction with Kroemer's contact characteristic, the general 
curve $j_c(E)$ may intersect the bulk velocity curve $v(E)$. 
The main difference between these two cases is that a Gunn effect 
mediated by charge dipole waves is seen only if $j_c(E)$ intersects 
the bulk velocity curve $v(E)$. 
If (\ref{ss4}) has a solution, 
Eq.~(\ref{ss2}) indicates that $E(y;J)$ 
approaches one of the solutions of (\ref{ss1}) as we leave the boundary
layer. The boundary layer at the receiving contact $y=1$
is a much narrower diffusive boundary layer of width $O(\epsilon\delta)$.
The field there is \cite{BonillaPD91}
\begin{eqnarray}
{\partial E\over\partial \eta} = \int_E^{E_{i}} v(E)\, dE ,\label{ss6}\\
{1-y\over\epsilon\delta} \equiv \eta = \int_{E_{L}}^{E(\eta)} 
{dF\over \int_{F}^{E_{i}} v(E)\, dE}\,, \label{ss7}\\ 
\mbox{where}\quad\quad \alpha_L \, (i_L + J) = -\frac{1}{\delta} 
                                       \int_{E_{L}}^{E_{i}} v(E)\, dE, 
\label{ss8}
\end{eqnarray}
($i=1,3$) whenever (\ref{ss8}) has a solution $E_L$. 

The idea now is to fix $J$ and to discuss for which values of $J$ 
the above construction yields a stationary solution. Additionally,
its stability will be considered. 
Clearly we may
distinguish different cases according to the values of 
the contact parameters ($i_i$, $\alpha_i$), $i=0,L$.
In what follows we shall assume for the sake of simplicity that the 
boundary layer equations at the receiving contact, Eqs. (\ref{ss7}) and
(\ref{ss8}), always have a solution,
and hence only the parameters of the cathode, ($i_0$, $\alpha_0$),
need to be considered.

The general situation encountered when constructing the stationary
solutions is the following. For each value of $J$, there are one 
or three values of the contact electric field at $x=0$, which are
solutions of Eq.\ (\ref{ss4}). We shall denote these field values 
by $E_{ci}(J)$, with $E_{c1}(J)<E_{c2}(J)<E_{c3}(J)$. 
Then the field profile in the injecting boundary layer is a 
monotonic solution of Eq.\ (\ref{ss2}), which joins $E_{ci}(J)$ 
($i=1,2,3$) to 
one of the solutions of $J=v(E)$ (outer solution). 
Furthermore, the outer solution may be a constant field profile
given by $E(y)=E_{l}(J)$ ($l=1,3$) which extends to
the end of the sample, where a narrow receiving boundary layer exists
[see Fig.\ \ref{sprof}]. 
For such an electric field profile, the bias is $\phi \approx E_{i}(J)$. 
The corresponding $J$-$\phi$ characteristics satisfies $J \approx v(\phi)$.
By this construction, we identify the portions of the
$J$-$\phi$ characteristics which follow the first
or the third branch of $v(E)$ (see the details below). Other
portions of the $J$-$\phi$ characteristics are flat, with $J=J_{f}$
for certain constant values of the current. 
The corresponding outer field profile is step-like with $E=E_2(J_f)
=E_{c2}(J_f)$ if $0<y<\Delta Y$ and $E=E_i(J_f)$ with $i=1,3$ 
if $\Delta Y<y<1$ [see Fig.\ \ref{sprof}]. $\Delta Y$ is chosen so as to
satisfy the bias condition $\phi \approx E_{2}(J_{f}) \Delta Y + (1-\Delta Y)
 E_{i}(J_{f})$. The flat part of the $J$-$\phi$ characteristics
corresponds to a bias range $E_{1}(J_f)<\phi<E_{3}(J_f)$ (see 
the details below). Finally, when $J$ is near the value $j^{sat}_0$, 
the field at the injecting contact is very large, the contact region
is almost depleted of electrons and its extension, $y =
\epsilon [E_{c3}(J)-E_i(J)]$ ($i=1,3$), may be comparable to
the length of the sample. Assuming the extension of the depletion layer
at the injecting contact is less than the sample length, the 
corresponding bias is $\phi \approx {1\over 2}\epsilon 
[E_{c3}(J)-E_i(J)]^2 + E_{i}(J)$ ($i=1,3$). 
The characteristics tends to saturate at $j_{0}^{sat}$. 
 
Following this general scheme, different possibilities may be 
distinguished according to the relative values of $j_{cm} < j_{cM} 
< j_{0}^{sat}$ with respect to $v_{m}<v_{M}$. Here, $j_{ck}=(1+\alpha_{0} 
i_{0}) v_{k}/(1+\alpha_{0} v_{k})$ and $v_{k}$, with $k=m,M$,  
refer to the minimum ($k=m$) and maximum ($k=M$) of the
contact, $j_{c}(E)$, and velocity, $v(E)$, curves, respectively .
We now discuss the different cases which appear for our 
velocity curve. 

\subsubsection{$j_{cm}<j_{cM}< j_{0}^{sat}<v_m<v_{M}$}
\begin{figure}     
\centerline{
    \epsfxsize=8 cm
    \epsffile{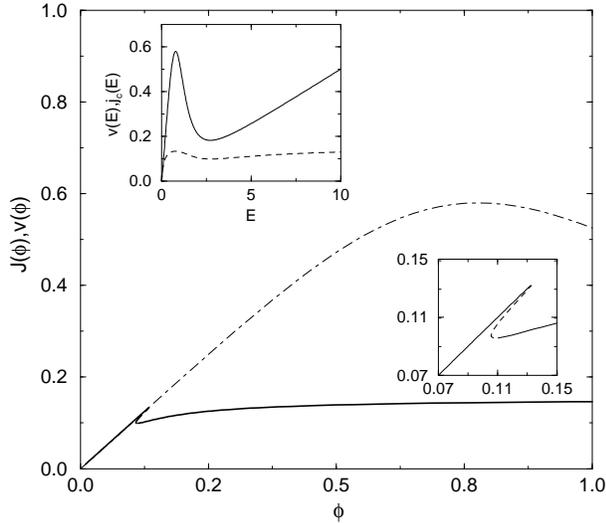}}
\caption{Stationary current-voltage characteristics, for $L=500$, 
$i_{0}=0.048$ and
$\alpha_{0}=9$, for which
$j_{cm}<j_{cM}<j_{0}^{sat}<v_{m}<v_{M}$, showing
bistability for biases $E_{1}(j_cm)<\phi<E_{1}(j_cM)$. The dashed 
line corresponds to the unstable solutions with $E(0;J) = E_{c2}(J)$.
For comparison the $v(\phi)$ curve is also plotted (dotted-dashed line). 
Insets: At the bottom, blowup of the bistable region. On top,
contact characteristics, $j_{c}(E)$, (dashed line) and 
velocity, $v(E)$, (continuous line) curves for this case.}
\label{stat1}
\end{figure}

In this case, see Fig.\ \ref{stat1} , we have for $0<J<J_{cM}$ 
a class of solutions joining $E_{c1}(J)$ and $E_{1}(J)$, with 
voltage $\phi \approx E_{1}(J)$.
For $0<\phi<E_{1}(j_{cM})$ the curve J-$\phi$ then follows the first
branch of $v(E)$. Furthermore, a second class of solutions 
joining $E_{c3}(J)$ and $E_{1}(J)$, will exist for $j_{cm}<J<j^{sat}_{0}$. 
In this case for $J$ not near $j^{sat}_{0}$, the voltage is given
by $\phi \approx E_{1}(J)$, and for $J$ near $j^{sat}_{0}$ it is by
$\phi \approx {1\over 2}\epsilon 
[E_{c3}(J)-E_1(J)]^2 + E_{1}(J)$. Then, in the characteristics the third 
branch starting at $\phi \approx E_{1}(j_{cm})$ follows the first branch
of $v(E)$ at the beginning, till it tends to saturate to $j_{0}^{sat}$
 for larger voltages (see Fig.\ \ref{stat1} ). Joining these two classes of 
solutions, there exists a third
class for $j_{cm}<J<j_{cM}$ which joins $E_{c2}(J)$ to $E_{1}(J)$,
with $\phi \approx E_{1}(J)$. These solutions are unstable,
and they give rise to the second branch in the characteristics,
Fig.\ \ref{stat1}, which also 
tend to follow the first
branch of $v(E)$. Note that for voltages $E_{1}(j_cm)<\phi<E_{1}(j_cM)$,
the two classes of stable stationary solutions coexist [see inset at
the bottom of Fig.\ \ref{stat1}]. Hysteresis between
them is then possible. 
\subsubsection{$j_{cm}<v_{m}<j_{cM}<j_{0}^{sat}<v_{M}$}
\begin{figure}   
\centerline{
    \epsfxsize=8 cm
    \epsffile{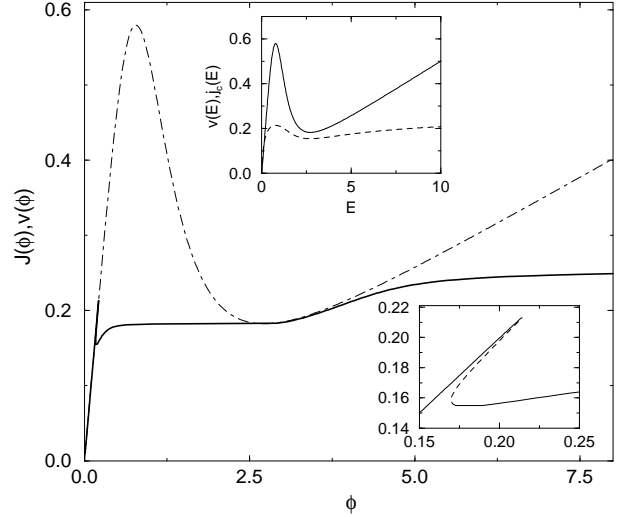}}
\caption{Stationary current-voltage characteristics, for $L=500$, 
$i_{0}=0.135$ and
$\alpha_{0}=8$, for which
$j_{cm}<v_{m}<j_{cM}<j_{0}^{sat}<v_{M}$, showing
bistability for biases $E_{1}(j_{cm})<\phi<E_{1}(j_{cM})$. The dashed 
line corresponds to the unstable solutions with $E(0;J) = E_{c2}(J)$. The
flat portion of the curve corresponds to $J=v_{m}$.
A Gunn effect mediated by moving depletion charge monopoles is
expected on a bias range which is a subinterval of $E_{M}<\phi_<E_{m}$.
For comparison the $v(\phi)$ curve is also plotted (dotted-dashed line). 
Insets: At the bottom, blowup of the bistable region. On top,
contact characteristics, $j_{c}(E)$, (dashed-line) and velocity, $v(E)$,
(continuous line) curves for this case. }
\label{stat2}
\end{figure}

In this case, see Fig.\ \ref{stat2} , the description is very similar to the
previous case, except on what concerns the third branch of the
$J$-$\phi$ characteristics. Now, this branch is composed of two types
of solutions: i) for $j_{cm}<J<v_{m}$ there is a class of solutions joining
$E_{c3}(J)$ and $E_{1}(J)$. Most of the time one has $\phi \approx E_{1}(J)$,
except for $J$ near $v_{m}$ that the solution is step-like
with $\phi \approx E_{m} \Delta Y + (1-\Delta Y) E_{1}(v_{m})$.
We expect that these solutions become unstable on a bias range
which is a subinterval of $E_{M}<\phi<E_{m}$ \cite{SIAM94,SIAM95}. Then a 
Gunn effect mediated by moving charge monopoles (which are charge depletion
layers) might appear (see companion paper \cite{companion}). 
For $v_{m}<J<j_0^{sat}$,
there is a class of solution joining $E_{c3}(J)$ and $E_{3}(J)$, with
$\phi \approx E_{3}(J)$ for $J$ not near $j_{0}^{sat}$ and
$\phi \approx{1\over 2}\epsilon 
[E_{c3}(J)-E_3(J)]^2 + E_{3}(J)$ for $J$ near $j_{0}^{sat}$. Then, the 
third branch of
the $J$-$\phi$ curve starts following the first branch of the $v(E)$
curve for $E_{1}(j_{cm})<\phi<E_{1}(v_{m})$, then it presents a flat
region for $E_{1}(v_{m})<\phi<E_{m}$ with $J=v_{m}$, 
corresponding to the step-like
solutions, and finally a region for $E_{m}<\phi$ which starts following
the third branch of the $v(E)$ curve and tends to saturate to $j_{0}^{sat}$
for larger voltages. Note the presence again of bistability for voltages
$E_{1}(j_{cm})<\phi<E_{1}(j_{cM})$ [inset at the bottom of Fig.\ \ref{stat2}] 
and hence the possibility of 
hysteresis. 

A similar situation to the one depicted above would appear for
$j_{cm}<j_{cM}<v_{m}< j_0^{sat}<v_{M}$.

\subsubsection{$v_{m}<j_{cm}<j_{cM}< j_0^{sat}<v_{M}$}
\begin{figure}   
\centerline{
    \epsfxsize=8 cm
    \epsffile{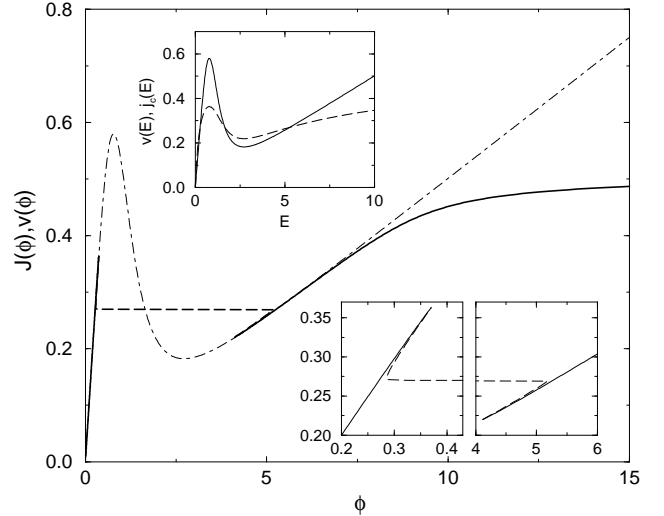}}
\caption{Stationary current-voltage characteristic for $L=500$, 
$i_{0}=0.27$ 
and $\alpha_{0}=4$, for which 
$v_{m}<j_{cm}<j_{cM}<j_{0}^{sat}<v_{M}$, 
showing the unstable stationary solutions 
(dashed line) with $E(0;J) = E_{c2}(J)$. The flat portion of
the curve corresponds to $J=i_{0}$. Note that no stable
stationary solution exists for $E_{1}(j_{cM})<\phi<E_{3}(j_{cm})$. Then
a Gunn effect mediated by moving charge dipoles is expected.
For comparison the $v(\phi)$ curve is also plotted (dotted-dashed line).
Insets: At the bottom, blowup of the unstable region. On top,
contact characteristics, $j_{c}(E)$, (dashed-line) and velocity, $v(E)$,
(continuous line) curves for this case.}
\label{stat3}
\end{figure}

The main difference for this case with respect to the previous ones,
relies on the second and third branches (see Fig.\ \ref{stat3}). Now, the 
third branch of the $J$-$\phi$ curve involves one single class of solutions 
joining $E_{c3}(J)$ and $E_{3}(J)$, with voltages $\phi \approx E_{3}(J)$
for $J$ not near $j_0^{sat}$, and $\phi \approx{1\over 2}\epsilon 
[E_{c3}(J)-E_3(J)]^2 + E_{3}(J)$ for $J$ near $j_{0}^{sat}$.
The second (unstable) branch is formed of two classes of solutions:
one class for $i_{0}<J<j_{cM}$ 
starting at $E_{c2}(J)$ and ending at $E_{1}(J)$, and another class
for $j_{cm}<J<i_{0}$, starting at $E_{c2}(J)$ and ending at $E_{3}(J)$.
For $J \approx i_{0}$, these solutions are step-like with the
voltage given through $\phi \approx E_{c2}(i_0) \Delta Y + 
(1-\Delta Y) E_{i} (J)$, with $i=1,3$ depending on the class of solutions
considered.
The (unstable) branch in the $J$-$\phi$ curve then starts following the 
first branch of the
$v(E)$ curve for $E_{1}(j_{cM})<\phi<E_{1}(i_{0})$, then it presents
a flat portion for $E_{1}(i_{0})<\phi<E_{3}(i_{0})$ with $J \approx i_{0}$,
and it ends following the third branch of the $v(E)$ curve for 
$E_{3}(j_{cm}) < \phi <E_{3}(i_{0})$. Note that
 for voltages
$E_{1}(j_{cM})<\phi<E_{3}(j_{cm})$ there is no stable 
stationary solution [see inset at the bottom of Fig.\ \ref{stat3}]. Thus we
 expect that the usual Gunn effect (mediated
by moving charge dipoles) will be present for these values of the bias
(see the companion paper \cite{companion}).

A similar situation appears for $v_{m}<j_{cm}<j_{cM}< v_{M}<j_0^{sat}$.

\subsubsection{$v_{m}<j_{cm}<v_{M}<j_{cM}< j_0^{sat}$}
\begin{figure}   
\centerline{
    \epsfxsize=8 cm
    \epsffile{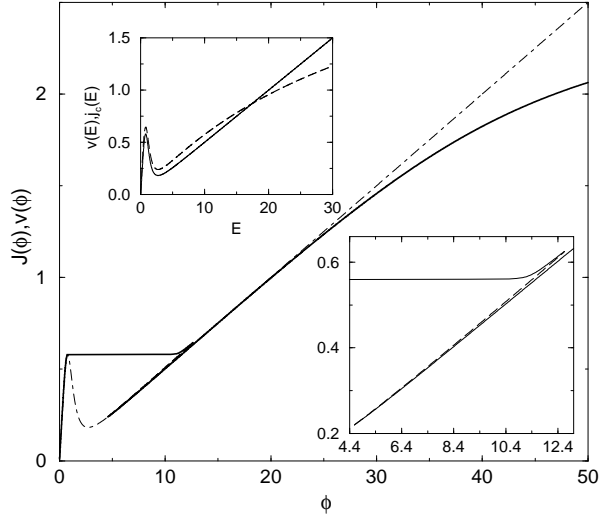}}
\caption{Stationary current-voltage characteristic for $L=500$, 
$i_{0}= 0.87$ 
and $\alpha_{0}=0.5$, for which
$v_{m}<j_{cm}<v_{M}<j_{cM}<j_{0}^{sat}$, 
showing bistability for biases $E_{3}(j_{cm})<\phi<E_{3}(j_{cM})$.  
The dashed line corresponds to the unstable stationary solutions 
with $E(0;J) = E_{c2}(J)$. 
The flat portion of the curve corresponds to $J=v_{m}$.
A Gunn effect mediated by moving depletion charge monopoles is
expected on a bias range which is a subinterval of $E_{M}<\phi<E_{m}$.
For comparison the $v(\phi)$ is also plotted (dotted-dashed line).
Insets: At the bottom, blowup of the bistable region. On top,
contact characteristics, $j_{c}(E)$, (dashed-line) and velocity, $v(E)$,
(continuous line) curves for this case.}
\label{stat4}
\end{figure}

In this case (see Fig.\ \ref{stat4}), the third branch of the $J$-$\phi$
characteristics is described as in the 
previous case. The first branch is composed of two types of solutions:
one class, for $0<J<v_{M}$, joining $E_{c1}(J)$ and $E_{1}(J)$, and
the other, for $v_{M}<J<j_{cM}$, joining $E_{c1}(J)$ and $E_{3}(J)$. For the 
first type of solutions, one has $\phi \approx E_{1}(J)$ except 
for $J \approx v_{M}$ that
$\phi= E_{M}\Delta Y + (1-\Delta Y) E_{3}(v_{M})$. These step-like
solutions are expected to become unstable in a subinterval of 
$E_M<\phi<E_m$ \cite{SIAM94,SIAM95}.
Then a Gunn effect mediated by moving charge monopoles (which are charge
accumulation layers) might appear (see the companion 
paper \cite{companion}). Thus, the first branch
starts following the first branch of the $v(E)$ curve for $0<\phi<E_{M}$,
then it presents a flat portion for $E_{M}<\phi<E_{3}(v_{M})$,
with $J=v_{M}$, and ends
following the third branch of the $v(E)$ for 
$E_{3}(v_{M})<\phi<E_{3}(j_{cM})$. The second (unstable) branch of the
$J$-$\phi$ curve is formed by a class of solutions that starts at 
$E_{c2}(J)$
and ends at $E_{3}(J)$, with $\phi \approx E_{3}(J)$. Then, this branch 
follows
the third branch of the $v(E)$ curve, for 
$E_{3}(j_{cm})<\phi<E_{3}(j_{cM})$. Note that for this range of bias, two
stationary stable solution coexists [see inset at the bottom of Fig.\ \ref{stat4}]
 and hysteresis may appear.  

\subsection{Phase diagram}
 
\begin{figure}   
\centerline{
    \epsfxsize=8 cm
    \epsffile{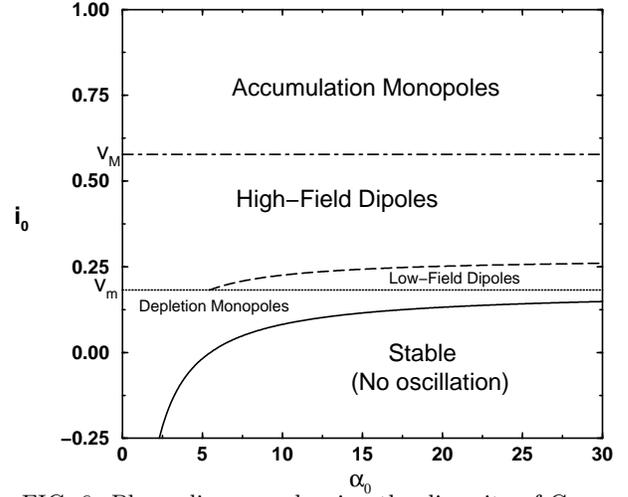}}
\caption{Phase diagram, showing the diversity of Gunn oscillations
that may appear depending on the values of the injecting contact
parameters, $i_{0}$ and $\alpha_{0}$.
The different separatrices correspond to $j_{0}^{sat}=v_{m}$ (continuous line), 
$j_{cm}=v_{m}$ (dotted line) and $j_{cM}=v_{M}$ (dotted-dashed line). Also
 depicted the separation
between low and high field charge dipoles, discussed in the companion
paper (dashed line).
}
\label{phdia}
\end{figure}

By collecting the information obtained in the previous subsections, the 
phase diagram describing the different behaviours of the system
can be sketched, in terms of the injecting contact parameters, 
$i_{0}$ and $\alpha_{0}$, Fig.\ \ref{phdia}.
Stable, non-oscillatory stationary solutions are expected for values of
$i_0$ and $\alpha_0$ such as $j_{0}^{sat} < v_m$, where 
$j_{0}^{sat}= \alpha_{0}^{-1} +i_{0}$. Otherwise, some kind of oscillatory
solution should be found.
Charge accumulation monopoles appear for $j_{cM}>v_{M}$ (or equivalently
for $i_{0}>v_{M}$), charge dipoles
for $v_{m}<j_{cm}<j_{cM}<v_{M}$, that is, for $v_{m}<i_{0}<v_{M}$,
and charge depletion monopoles for $j_{0}^{sat}>v_{m}$ with $j_{cm}<v_{m}$ 
($i_{0}<v_m$).
For completeness, also the separation between
low field and high field dipoles, discussed in the companion 
paper \cite{companion}, has been depicted.
It is worth noting that this rich phenomenology of oscillatory states appears just
by changing the value of the contact parameters. This 
fact should be taken
into account in analyzing the Gunn effect in real systems, where, as 
mentioned
before, a wide range of values for the contact parameters, depending on the
metal used and preparation procedures, may appear.

\section{Discussion} 
We have presented a general formulation for the derivation of the
boundary conditions imposed by metal-semiconductor contacts on 
semiconductor systems. According to this general formulation,
the appropriate boundary conditions for ideal metal-semiconductor
contacts are linear relations between 
the normal derivative of the electric field at the contacts and
the electron current there. For the classical unipolar drift-diffusion 
Kroemer's model of the Gunn effect, these boundary
conditions are of mixed type. In this paper, we have investigated how 
the boundary conditions for ideal metal-semiconductor contacts affect 
the stationary solutions of the Kroemer model, and their
stability. Depending on the values of the contact parameters,
bistability and hysteresis may appear. Moreover, for some range of
parameters no stable stationary solution is expected to occur. In those
parameter ranges we expect to find the Gunn effect. Numerical simulations
show that different types of Gunn effect appear, mediated by a 
variety of waves: (i) charge monopole accumulation wavefronts,
(ii) monopole depletion wavefronts,  or (iii) charge  dipole waves
(high and low electric field domains). Why these types of Gunn
effect appear in the simulations will be explained by the asymptotic
theory of the companion paper \cite{companion}. It suffices to say that without
 this 
theory we would have missed significant possible instabilities.
For example (ii) seems to have been missed by earlier workers, in 
spite of past extensive simulations of Kroemer's model \cite{Grubin1}.
With our boundary conditions, the previously described types of Gunn effect 
are found in the following ranges of dimensionless critical contact currents: 
(i) $j_{cM}>v_M$ (ii) $j_{cm}<v_{m}$ and $j_0^{sat}>v_{m}$, and (iii) 
$v_m<j_{cm}<j_{cM}<v_M$. Here $j_{cM}=j_c(E_M)$, $j_{cm}=j_c(E_m)$, 
$j_0^{sat}$ are the critical currents, and $v_m$ and $v_M$ are the minimum and 
maximum values of the electron drift velocity $v(E),\, E>0$. 
When we want to characterize experimental samples displaying the
Gunn effect, it is important to bear in mind the great influence of the 
contact parameters on the type of wave mediating the Gunn effect. A
wide range of values for these contact parameters may be obtained depending 
on the type of metal used or on the contact preparation procedure followed.

N-shaped velocity curves occur naturally in recently observed self-sustained
oscillations in weakly-coupled n-doped GaAs/AlAs superlattices (see Ref.\
\onlinecite{kastrup97} for the most complete data so far). In these
superlattices there is strong indirect evidence for a Gunn effect mediated
by charge accumulation monopoles through photocurrent and photoluminescence
measurements \cite{ICPF94}. It is hard to say at this point which other 
possibilities
of those found in our analysis might be realizable in these systems. An
important issue to be decided is the form of the boundary conditions.
Our analysis needs to be modified in order to be extended to these systems,
as quantum tunneling plays an essential role in the injection of carriers 
through contact regions. 

The most used velocity
curves $v(E)$ for the classical Gunn effect in bulk n-GaAs lack the
third branch after $v_m$. The reason is that avalanche breakdown appears
at electric fields smaller than $E_m$. The avalanche field is smaller
for the longer samples needed to observe the Gunn effect and this 
precludes reaching the high fields on the third branch of $v(E)$. 
Then low field dipole domains and charge depletion monopoles are not
observed in the usual bulk samples or in strongly coupled 
superlattices with wide minibands, which are analogous to them
\cite{buttiker}.

\section{ACKNOWLEDGMENTS}
\label{acknowledgements} 
We thank Dr.\ M.\ Bergmann for fruitful conversations, and acknowledge 
financial support from the the Spanish DGICYT through grants PB94-0375
and PB95-0881, and from the EC Human Capital and Mobility Programme 
contract ERBCHRXCT930413. One of us (G.G.) acknowledge support by the 
Generalitat de Catalunya.

\appendix

\setcounter{equation}{0}

\section{Derivation of Eq.~(\ref{TE}) by means of SRH statistics}
\label{appendixA}
Let us consider the elementary kinetic process: 
$q^{m} \rightleftharpoons q^{n}$ describing the charge transport
through the junction. By assuming the validity of the SRH statistics
to describe this process, the following expression for its kinetic
rate, $J_{n}$, can be obtained~\cite{Vanvliet}:
\begin{eqnarray}
J_{n} &=& \int \mbox{d} E_{n} \int \mbox{d} E_{m} D_{m}(E_{m}) 
     D_{n}(E_{n}) \nonumber \\
       & & \times [ f_{m} (E_{m}) (1- f_{n} (E_{n})) 
     \gamma_{mn} (E_{m}, E_{n}) \nonumber \\
       & & - f_{n} (E_{n}) (1 - f_{m} (E_{m})) 
       \gamma_{nm} (E_{m}, E_{n})] ,
\label{rate0}
\end{eqnarray}
where $D_{a}(E_{a})$, $a=n,m$, is the density of states of system $a$,
$f_{a} (E_{a})$ its occupation function, given through the
FD distribution $f_{a} (E_{a}) = (1 + 
e^{\beta(E_{a} - E_{F_{a}})})^{-1}$, with $E_{F,a}$ the corresponding
Fermi level. 
$\gamma_{mn} (E_{m}, E_{n})$ [resp. $\gamma_{nm} (E_{m}, E_{n})$]
is the probability per unit time
for the transition between states of energy $E_{m}$ and $E_{n}$
[resp. $E_{n}$ and $E_{m}$].
At equilibrium, we must have $J_{n} = 0$ and $F_{m} = F_{n}$,
with $F_{a} = E_{F,a} - e V_{a}$ being the corresponding
quasi-Fermi levels. This implies $E_{F,n} - E_{F,m} = e\, (V_n - V_m) = 0$
(using the assumed continuity of the electric potential at the contact). 
These equations follow from Eq.~(\ref{rate0}) if the latter is supplemented 
with the following detailed balance relation:
\begin{equation}
\gamma_{mn} (E_{m}, E_{n}) = 
\gamma_{nm} (E_{m}, E_{n}) e^{\beta (E_{m} - E_{n})} .
\label{detailed}
\end{equation}
A term $\beta\, (V_n - V_m)$ has to be added to the argument of the exponential
in (\ref{detailed}) if $V_n\neq V_m$; see \cite{Noltros3} for a more general 
case. We now substitute Eq.~(\ref{detailed}) into Eq.~(\ref{rate0}) and use 
the equations 
$$ E_{C} = E_{C}^{0} - e V_{n},\quad\quad\quad e\phi_b^0 = E_C - F_m$$ 
($E_{C}^{0}$ is bottom of the semiconductor conduction band and $e\phi_b^0$ 
is the height of the contact barrier). After straightforward manipulations, 
we derive (\ref{TE}), in which the transition coefficient $\lambda_{0}$ is:
\begin{eqnarray}
\lambda_{0} &=& \int \mbox{d} E_{n} \int \mbox{d} E_{m} D_{n}(E_{n})
                D_{m}(E_{m}) (1 - f_{n} (E_{n})) \nonumber \\
             & & \times (1 - f_{m} (E_{m}))
 \gamma_{nm} (E_{m}, E_{n}) e^{\beta (E_{C}^{0} - E_{n})} .
\label{lambda}
\end{eqnarray}
When the semiconductor is non-degenerate, we may approximate $1 - 
f_{n}(E_{n}) \approx 1$, whereas for a metal we may approximate $f_{m}(E_{m})$ 
by its equilibrium value. Then, for this case,
 $\lambda_{0}$ is a function of $T$ only.

\end{multicols}

\end{document}